\def\lsim{\mathrel{\rlap{\lower4pt\hbox{\hskip1pt$\sim$}}
    \raise1pt\hbox{$<$}}}         
\def\gsim{\mathrel{\rlap{\lower4pt\hbox{\hskip1pt$\sim$}}
    \raise1pt\hbox{$>$}}}         

\def\ie{\hbox{\it i.e. }}
\def\etc{\hbox{\it etc... }}
\def\eg{\hbox{\it e.g. }}
                                                     
\def\etal{\hbox{\it et al. }}

\def\g2{ GeV$^2$}
\def\xi2{$\chi^2_{d.o.f}$}

\documentstyle[12pt,epsfig]{article}
\begin{document}
\begin{titlepage}

\rightline{LYCEN 9896 (November 1998)}

\vskip 0.8cm
\centerline {\bf APPROXIMATE SOLUTION OF THE BFKL EQUATION} 
\centerline {\bf APPLICABLE AT HERA}
\vskip 2cm
\centerline{\bf P. Desgrolard ({\footnote{E-mail: desgrolard@ipnl.in2p3.fr}}),
L. Jenkovszky ({ \footnote{E-mail: jenk@bitp.kiev.ua}}), 
F. Paccanoni ({ \footnote{E-mail: paccanoni@pd.infn.it}   })} 
\vskip 1cm
($^1$) {\it Institut de Physique Nucl\'eaire de Lyon, IN2P3-CNRS et
Universit\'e Claude Bernard, F69622 Villeurbanne Cedex, France.}

($^2$) {\it Bogolyubov Institute for Theoretical Physics,
National Academy of Sciences of Ukraine, Kiev-143,
Ukraine.}  

($^3$) {\it Dipartimento di Fisica, Universit\'a di Padova,
Istituto Nazionale di Fisica Nucleare, Sezione di Padova, 
via F.Marzolo, I-35131 Padova, Italy.}

\vskip 3cm
\centerline {\bf Abstract}
\vskip 1cm

We suggest a formula interpolating between the known asymptotic
regimes of the BFKL equation as the approximate solution of that equation.
The parameters appearing in this interpolation are fitted to the data on 
deep inelastic scattering in a wide range of the kinematical variables.
Care is taken of the large-$x$ domain as well, outside the HERA kinematical 
region. The boundaries and the interface between various dynamical regimes 
are also studied.

\end{titlepage}

\bigskip

{\bf 1 INTRODUCTION }

\medskip
The Balitski-Fadin-Kuraev-Lipatov (BFKL) equation \cite{bfkl} is a 
relativistic bound state equation for two reggeized gluons. Although first 
applications concerned $\gamma \gamma$ scattering, the equation originally
was derived for the hadronic (on mass shell) scattering amplitude in quantum 
chromodynamics (QCD) at high energies $\sqrt s$ and fixed momentum 
transfer $\sqrt{-t}$ in the leading logarithmic approximation (LLA), implying 
the collection of all terms of the type $(\alpha_s \ln s)^n$, $\alpha_s$ being
the QCD ("running") coupling constant. In this approximation, the total 
cross section increases rapidly, as 
$$\sigma_{tot}^{LLA}\sim {s^\omega\over{\sqrt{\ln s}}},           \eqno(1.1) $$
where $\omega$ is (in the LLA) the position of the rightmost singularity  in 
the complex angular momentum plane of the $t$-channel partial wave with the 
vacuum quantum numbers (Pomeranchuk singularity), given by
$$\omega={g^2\over{\pi ^2}}N\ell n 2                              \eqno(1.2) $$
for the gauge group $SU(N)$ (N=3 in QCD), with a gauge coupling constant 
$g=\sqrt{4\pi \alpha_s}$.

Simply stated, the hope is that by solving the BFKL equation, one will be 
able to restrict the freedom available in the Regge pole theory, namely to
determine the form of the leading singularities in the angular momentum 
plane, the form and the values of the parameters of the vacuum (Pomeranchuk) 
trajectory \etc Of all these ambitious expectations only (1.1) was of practical use,
although still subject of uncertainties coming from the convergence condition
$\alpha_s\ln s \le 1$ and kinematical limitations. Any extrapolation of 
$\alpha_s$ to $t=0$ should have discouraged the direct application of (1.1) 
to fit measured hadronic cross sections. Really, even with a conservative value 
of $\omega\ge 1.3$, optimistic attempts in fitting data in the late 70-ies and 
early 80-ies were soon abandoned.

With the appearence of large virtualities, a new "hard" scale at HERA offered 
new possibilities for the interpretation of the "QCD-Pomeron"~: most of the 
HERA results were claimed to confirm the existence of a "hard" Pomeron, 
manifest in the rapid rise of the structure functions, photoproduction of 
heavy vector mesons \etc While the experimental results were claimed to 
confirm quantitatively the predictions of the BFKL equation (rapid rise in $s$ 
or in $x^{-1}=s/Q^2$) and of the perturbative QCD in the whole, the 
interpretation proved once again to be misleading. Actually, the 
corrections to the LLA in the next-to-leading logaritmic approximaton (NLLA) 
were found large and distructive, lowering substantially the value of the 
Pomeron intercept \cite{fali} and raising doubts about the "hardness"  
of diffraction.
  
What remains now from the "QCD Pomeron" is that it has a complicated 
$j$-plane structure and the rightmost (leading) singularity is located 
somewhere above unity. Since there is little hope that the whole perturbative 
series will ever be summed, and waiting for possible numerical solutions of the 
BFKL equation, all we can do now is to start from a "supercritical" 
($\alpha(0)> 1$) Pomeron and adjust it to the data.

An important point to be mentioned here is that the above Pomeron is meant 
at the Born level, \ie it should be subject to a subsequent unitarization 
procedure. Such a procedure is very complicated already in case of hadronic
(on mass shell) reactions and becomes even more tricky as $Q^2\neq 0$. A
practical way out from this situation may be along the lines suggested in 
\cite{ckmt}, namely by introducing a $Q^2$-dependent "effective" trajectory 
of a single, simple, factorizable supercritical Pomeron pole, absorbing 
the effect of its complicated $j$-plane structure as well as those from the 
possible unitarity effects (that cannot be calculated exactly anyway) in the 
form of its "effective" trajectory. The free parameters are then adjusted to the 
experimental data. All what remains now from QCD is that $\alpha(0) > 1$,
which reflects the present status of the solutions of the BFKL equation at 
finite $Q^2$.
 
However, the solution of the BFKL equation is known exactly in the 
asymptotic, $Q^2\to\infty$ limit as well, namely
$$F_2(x,Q^2)\sim\exp{\sqrt{\gamma_1\ell n(1/x)\ell n{\ell n Q^2}}}. \eqno(1.3)$$
We suggest to use the two known solutions of the BFKL equation, (1.1) and (1.3)
as boundary conditions and interpolate between these two by employing a 
minimal number of additional parameters. In our opinion this is the simplest 
solution of the problem, yet applicable to realistic processes at arbitrary 
values of $Q^2$. By fitting the parameters to the experimental data, we hope 
to be consistent with unitarity and find the boundaries of different dynamical 
regimes, namely those governed by the BFKL equation or QCD evolution.

In our previous paper \cite{djp98} we have already investigated such an 
interpolation within the GLAP evolution equation \cite{glap}
by assuming the "BFKL Pomeron" (1.1) to be the 
input, subject of a subsequent evolution. Actually, the asymptotic solutions 
of both the BFKL and GLAP equations have the same form (1.3) but the relevant 
paths may be different. 
Aiming at a high quality and reliable fit at small $x$, we have substanially
improved as compare to our previous paper \cite{djp98} the quality and range 
of the fits in the ("subsidiary") large-x domain. Still, the real paths of the 
solutions remain ambiguous for two main reasons: uncertainties in the low 
$Q^2$, nonperturbative behavior (remaining outside of both the GLAP and 
BFKL equations) and unitarization (to effect both solutions), whose 
role increases with increasing $Q^2$ and decreasing $x$. Further work in 
this direction is needed.

Our paper is organized as follows. In Sect. 2 we present our interpolating 
solution, in Sec. 3 we find the value of the adjustable parameters by fitting 
to the data on deep inelastic scattering; our conclusions are given 
in Sect. 4.

\bigskip

{\bf 2 THE MODEL}

\medskip

We use the standard kinematic variables to describe deep
inelastic scattering (DIS)~:
$$ e(k) \ +\ p(P) \ \to\ e(k')\ +\ X\ ,                          \eqno(2.1)$$
where $ k, k', P$ are the  four-momenta of the incident electron, scattered
electron and incident proton.
$ Q^2$ is the negative squared four-momentum transfer carried by
the virtual exchanged photon (virtuality)
$$ Q^2 \ =\ -q^2\ =\ -(k-k')^2\ .                                \eqno(2.2)$$
$x$ is the Bj\"orken variable
$$ x\ =\ {Q^2\over 2P.q} \ .                                     \eqno(2.3)$$
$W$ is the center of mass energy of the ($\gamma^*, p$) system, related to the 
above variables by 
$$ W^2 \ =\ Q^2{1-x\over x}+m_p^2 \ ,                     \eqno(2.4)$$ 
with $m_p$ being the proton mass. 

\medskip

{\bf 2.1 Structure function for low $x$ and all $Q^2$}

\medskip

According to \cite{djp98}, we adopt the
following ansatz for the small-$x$ singlet part (labelled by the
upper index $S,0$) of the proton structure function, interpolating
between the "soft" (1.1) and "hard" (1.3) regimes           

$$F_{2}^{(S,0)}(x,Q^2) =  \ G(Q^2)\     
e^{\Delta (x,Q^2)}  ,                                             \eqno(2.5)$$
with 
$$ G(Q^2)=
A\ \left({Q^2\over Q^2+a}\right)^{1+\widetilde{\Delta} (Q^2)}\ ,  \eqno(2.6)$$
$$ \widetilde{\Delta }(Q^2) =\epsilon+\gamma_1\ell n {
\left(1+\gamma_2\ell n{\left[1+{Q^2\over Q^2_0}\right]}\right)} ,
                                                                  \eqno(2.7)$$
and
$$\Delta (x,Q^2) = \left(\widetilde{\Delta } (Q^2)\  \ell n{1\over x}\right)
                   ^{f(Q^2)},                                     \eqno(2.8)$$
$$ f(Q^2) = {1\over 2}\left( {1+e^{-{Q^2/Q_1^2}}}\right) .        \eqno(2.9)$$
The function $f(Q^2)$ has been introduced in order to provide for the
transition from the Regge behavior, where $f(Q^2)=1$, to the 
asymptotic solution (1.3) of the BFKL evolution equation where                      
$f(Q^2)=1/2$. Alternative choices for this function, satisfying the
boundary conditions, cannot be 
excluded, but we find our way of interpolation via (2.9) to be the simplest
possible.

It is customary to define an "effective" Pomeron intercept 
$\alpha_{\cal P}^{eff}$, which is in gereral $x-$ and $Q^2-$ dependent,
by rewriting the proton SF, introducing an "effective power" 
$\Delta^{eff}(x,Q^2)$
$$F_{2}^{(S,0)}(x,Q^2) = G_1(Q^2)\ 
x^{-\Delta^{eff}(x,Q^2)}\ ,                                     \eqno(2.10) $$
where the two effective quantities satisfy 
$$\Delta^{eff}(x,Q^2)=\alpha_{\cal P}^{eff}(x,Q^2)-1\ .         \eqno(2.11)$$
This definition with our parametrization leads to the identification
$$ G_1(Q^2)\ =\ G(Q^2) \ ,\qquad
\Delta^{eff} (x,Q^2)= {\Delta(x,Q^2)\over\ell n{1\over x}}\ .    \eqno(2.12)$$
It is worth noting that, in the limit
$$f(Q^2)\simeq 1\ ,\quad {\rm \ie when}\ Q^2\ll Q_1^2 \ ,        \eqno(2.13)$$
the proton singlet (Pomeron component) SF reduces to 
$$F_{2}^{(S,0)}(x,Q^2\ll Q_1^2) \ \simeq\
G(Q^2) \ x^{-\widetilde{\Delta}(Q^2)}\ .                     \eqno(2.14)$$
We recover the standard (Pomeron-dominated) Regge behavior (with a 
$Q^2$-dependence in the effective Pomeron intercept).
Consequently, within this approximation
$$\Delta^{eff} (x,Q^2)\ \simeq\ {\widetilde{\Delta}(Q^2)}\ .  \eqno(2.15)  $$
Therefore, at small and moderate values of $Q^2$ (to be specified from the fits
(see below), the exponent ${\widetilde{\Delta}(Q^2)}$ may be interpreted as a 
$Q^2-$dependent (and of course $x-$independent) effective power.

\medskip

By construction, the model (singlet component) has the following $Q^2$ 
limits~:

\smallskip                                                                  

- a) $Q^2\to\infty$, fixed $x$:
$$ F_{2}^{(S,0)}(x,Q^2\to \infty)\to A\
\exp\left({\sqrt{\gamma_1\ell n\ell n{Q^2\over Q_0^2}\                        
\ell n{1\over x}}}\ \right)\ ,                                    \eqno(2.16)$$
which is the asymptotic solution of the BFKL and GLAP evolution equation 
(see Sect.1).

\smallskip

- b) $Q^2\to 0$:
$$F_{2}^{(S,0)}(x,Q^2\to 0) \to A\
e^{\Delta (x,Q^2\to 0)} \
\left({Q^2\over a}\right)^{1+\widetilde{\Delta }(Q^2\to 0)}       \eqno(2.17)$$
with $$ \widetilde{\Delta }(Q^2\to 0) \to
\epsilon+\gamma_1 \gamma_2   { \left({{Q^2\over Q^2_0} }\right)}\
\to\ \epsilon ,                                                   \eqno(2.18)$$
$$ f(Q^2\to 0) \to 1 ,                                            \eqno(2.19)$$
whence
$$F_{2}^{(S,0)}(x,Q^2\to 0) \to A\ \left( {1\over x}
\right)^\epsilon \ \left({Q^2\over a}\right)^{1+\epsilon} \ \propto
Q^2 \ \to 0\ ,                                                    \eqno(2.20)$$
as required by gauge invariance.

Apart from the (singlet, or Pomeron) component, a non-singlet component 
(sub-leading, or secondary Reggeons in terms of the Regge pole model) is 
also present at small $x$. Their contribution will be lumped in an 
"effective Reggeon" term, labelled by $(NS,0)$ with the $\rho, \ \omega, 
\ f$ and $A_2$ reggeons absorbed in an effective trajectory with an intercept 
$\alpha_r$, namely:  
\medskip

$$F_{2}^{(NS,0)}(x,Q^2) =\ H(Q^2)\ x^{1-\alpha_r}\ ,\quad 
H(Q^2)= \ B \left({Q^2\over Q^2+b}\right)^{\alpha_r} \        . \eqno(2.21)  $$
The resulting low-$x$ proton SF becomes
$$F_{2}^{(0)}(x,Q^2) =\ 
F_{2}^{(S,0)}(x,Q^2) +\ F_{2}^{(NS,0)}(x,Q^2)\ .                \eqno(2.22)   $$

\medskip

{\bf 2.2 Extension to large $x$}

\medskip

Our aim is to fix the free parameters appearing in our approximate solution of 
the BFKL equation (2.22). To this end, we could, in principle, fit (2.22)
to the small-$x$ data, where other contributions are negligeable. Whatever 
attractive, such a straighforward approach is not feasible since the small 
$x$ domain where contributions to the SF other than (2.5) may be neglected is 
very narrow, making any fit unreliable. Moreover, the relevant limits depend 
on the models used, as shown explicitely in \cite{pakis}.
Therefore we extend our model (2.22) by a "large-$x"$ part in order to have 
reasonable fits in a wide range of $x$. 

In our previous paper \cite{djp98} the relatively simple and efficient 
model of Capella \etal\cite{ckmt} (CKMT) was used for that purpose. Below,
aiming a better fit to the large-$x$ data, we extend that model, by 
introducing additional adjustable parameters as follows \cite{dlm98}     
$$F_{2}(x,Q^2) =\                                                   
F_{2}^{(S,0)}(x,Q^2) \cdot (1-x)^{P(Q^2)} +\ 
F_{2}^{(NS,0)}(x,Q^2)\cdot (1-x)^{R(Q^2)}\ .                  \eqno(2.23)   $$
Similar to  \cite{dlm98}, we use the following $Q^2$-dependent exponents 
of the large $x$ factors (with 6 additional parameters)
$$ P(Q^2)=\ p_\infty+{p_0-p_\infty\over 1+Q^2/Q^2_p}\ ,  
\quad R(Q^2)=\ r_\infty+{r_0-r_\infty\over 1+Q^2/Q^2_r}\ .   \eqno(2.24)$$  
    
\medskip

{\bf 2.3 Total cross-section for $(\gamma^*,p)$ scattering}

\medskip
 
The structure function is related to the  total
cross-section of virtual Compton scattering (or approximatively to the 
transverse cross-section, if the longitudinal component is neglected) by
$$\sigma _{tot}^{\gamma^*,p} (x,Q^2)=
{4\pi^2\alpha\over Q^2}{1\over 1-x}\left(1+{4m^2_p x^2\over Q^2}\right)\
F_{2}(x,Q^2)  \ =\  \sigma _{tot}^{\gamma^*,p} (W,Q^2) ,\       \eqno(2.25)$$
where $x$ is replaced by using (2.4).

In the limit $Q^2\to 0$, the SF vanishes as $Q^2$
and we write the total cross-section for $(\gamma,p)$ scattering
(with real photons), as a function of the 
center  of mass energy $W$ 
$$\sigma _{tot}^{\gamma,p} (W)=
4\pi^2\alpha\ {\rm lim}_{Q^2\rightarrow 0}
\left[{F_2(x=Q^2/W'^{2},Q^2)\over Q^2}\right] $$
$$= 4\pi^2\alpha\ \left( A\  a^{-1-\epsilon}\ 
W'^{2\epsilon} + B\ b^{-\alpha_r}\  W'^{2(\alpha_r-1)}\ \right) \ ,
                                                                \eqno(2.26) $$
with $W'^2=W^2-m^2_p$.
\bigskip

{\bf 3 FITTING TO THE DATA; RESULTS}
                                             
\medskip

In our fitting procedure the experimental data sets from "H1" 
\cite{ah95,ai96,ad97}, "ZEUS" \cite{de96,br97,br98}, "E665" \cite{ad96}, 
"NMC" \cite{ar95}, SLAC\cite{slac}, "BCDMS" \cite{be89}
for the proton structure function $F_{2}(x,Q^2)$ were used as well as data
points \cite{ca75} on the $(\gamma,p)$  total cross-section
$ \sigma_{tot}^{(\gamma,p)} (W)$, 
in the kinematical region with $0\leq Q^2 \leq 5000$ \g2,
$0<x\leq 0.75$, $W\geq 3$ GeV. The total number of data for this
"complete" ensemble is 1253 (see Table 1).

The resulting fits may be commented as follows.

The use in \cite{djp98} of an "economic" (only 8 free parameters) large-$x$ 
extension {\it a la} CKMT\cite{ckmt} resulted in good fits (\xi2 $\sim$ 0.67) 
within a rather limited domain in $x\leq 0.1$, with a selection of about 310 
data points (mainly those of the low $x$ H1 data). 

Better fits at large $x$ (including the BCDMS and 
SLAC data) can be achieved (at the price of 6 additional parameters with the 
large $x$ extension, presented above. 
Namely, we obtained 
\xi2 $\sim$ 1.16, distributed among each subset of 
data as shown in Table 1.                           
This is a considerable improvement with respect to the small $x$, all $Q^2$ 
results of \cite{djp98}. Interestingly, the present improvement of the 
large-$x$ behavior has little effect on the values of the parameters 
governing the low $x$ behavior and fitted previously \cite{djp98}.
As noted, 6 new parameters were introduced in our present version of the 
model (see Table 2). Among a total of 16 parameters, {\bf 14} are free 
(instead of 8 in \cite{djp98}), the remaining 2 being fixed in the following 
way:

1. Similar to \cite{djp98}, we choose the "canonical" value \cite{eps}
 $\epsilon=0.08$;

2. Similar to \cite{djp98}, we estimate from QCD the parameter
$\gamma_1=16N_c/(11-2f/3)$ with four flavours $(f=4)$ and
three colors $(N_c=3)$, it equals 5.76. It corresponds to the
asymptotic regime (when $Q^2\rightarrow\infty$, or $f(Q^2)\rightarrow 1/2))$,
far away from the region of the fits, where  $f=1$, 
hence the value ${\gamma_1}=\sqrt{5.76}=2.4$ 
is more appropriate in the domain under consideration. 
Fits with $\gamma_1$ free confirm this choice.

To compare with, the proton structure function and $(\gamma^*,p)$ cross
section are calculated in the ALLM model \cite{allm} in  
the whole experimentally investigated kinematical range. The ALLM fits to the 
data are good (\xi2 $\sim$ 1.1, recalculated) with a total of 23 
adjustable parameters.
This is also to be compared with the 21 parameters used in \cite{dlm98}, where 
a quite good fit (with only a slightly better \xi2) is also presented.  

The results of our fits for the structure function versus $x$
for fixed selected $Q^2$ are shown in Figs.~1-2.

The total cross section for real photons on protons as function 
of $W^2$ is displayed in Fig. 3.

The agrement with the data is impressive except may be for the SF reported
at the highest $Q^2$-value (=5000 \g2). We recall once again that this result 
is obtained with 14 free parameters.

\medskip

\noindent
 {\bf $Q-$slope as a function of $x$}
\medskip

Known as a useful tool in the sudies of low $x$ region, the derivative of the 
SF with respect to $\ell n Q^2$ 
$$ B_{Q}(x,Q^2)={\partial F_{2}(x,Q^2)\over \partial (\ell n Q^2) } \eqno(3.1)$$
($Q-$slope for brevity) measures
the amount of the scaling violation and eventually shows the
transition from soft to hard dynamics. 
It was recently extracted from the HERA data \cite{br98}, where the variables 
$x$ and $Q^2$ are strongly correlated, because
it is implied that, for a limited acceptance (as it is the case       
in the HERA experiments) and for a fixed energy, one always
has a limited band in $Q^2$ at any given $x$, with average $<Q^2>$
becoming smaller for smaller $x$.
 
From the theoretical point of view however, this derivative depends on two
variables ($x$ and $Q^2$) which are
quite independent and one is not restricted to
follow a particular path on the surface representing the $Q-$slope.
That is why we discussed in \cite{djp98} the Q-slope as a three-dimensional
quantity.
Here we adopt a more pragmatic attitude and we only compare in Fig.~4 the
predictions of our model with the available experimental results for the
special experimental set of ($x,Q^2$) chosen in \cite{br98}.
Our predictions are quite in agreement with the data,
as it may be expected from a good fit of the structure function implying 
also agreement with its experimental derivatives ~:
an increasing $Q-$slope when $x$ decreases down to $\sim 10^{-4}$, then a 
"turn over"~: the slope decreasing at lower $x$ (or $Q^2$) values. 
Notice that the turn over point, located at $Q^2\sim 2$ GeV$^2$, may be related 
to the different behaviors of the gluon and $q\bar q$ sea distributions. 

\medskip

\noindent
 {\bf $x-$slope as a function of $Q^2$}

\medskip

The derivative of the logarithm of the SF with respect to $\ell n 1/x$
$$B_x(x,Q^2)={\partial\ell n F_{2}(x,Q^2)\over\partial(\ell n(1/x))}\eqno(3.2)$$
($x-$slope for brevity), when measured in the Regge region,
can be related (for low $x$) to the Pomeron intercept. In Fig.~5, we 
compare the calculated $x-$slope with the quantity called "effective power" 
($\lambda_{eff}$ ).
In fact, the measured ($x-$dependent) quantity should not be confused with 
the "effective power" introduced in (2.10-12) ($\Delta^{eff}(x,Q^2)$ in our 
notation), it is really the $x-$slope.
The agreement between experiment and theory is very good.

Finally, in Fig.~6
the $Q^2$-dependence of this calculated derivative together with the 
power $\tilde\Delta$ (2.8) is shown for some low 
$x$ - values. On the same figure, the behavior of the
function $f(Q^2)$ (2.9) is also shown. In our model, Regge pole                 
behavior is equivalent to the condition that $f(Q^2)$ is close to
unity
$$f(Q^2)\approx 1 \ ;\quad
{\partial\ell nF_{2}\over\partial(\ell n(1/x))}\ \approx\Delta^{eff}\ 
\approx\tilde\Delta  \  .                                           \eqno(3.3)$$
This lower limit, marked on Fig.~6 (tentatively approximated           
within a 2 \% accuracy for the function $f(Q^2)$), is 
located near  $60$~\g2 .
Until this landmark, the power $\tilde\Delta$ indeed 
remains very close to the $x-$slope. 
Beyond, Regge pole behavior is not valid (since $f\ \ne 1$) and $\tilde\Delta$ 
cannot be considered as the $x-$slope any more. 
On the other hand, the $x-$slope turns down as $Q^2$ increases, approaching 
its "initial value" of $\approx 0.1$ at largest $Q^2$ and coming 
closer to the unitarity bound. Notably, at large $Q^2$ the derivative 
gets smaller as $x$ decreases, contrary to the general belief 
that dynamics becomes harder for smaller $x$, but in accord with an 
observation made in \cite{dero}. Care should be however taken in 
interpreting the "hardness" of the effective power outside the Regge 
region.

The transition region occurs when $f(Q^2)$ goes from 1 to 1/2 \ie in a band in
$Q^2$ estimated between $\sim 60$~\g2 and $\sim 5000$~\g2.

\bigskip

{\bf 4 CONCLUSIONS}
                                             
\medskip
                     
The consequences of the present studies are manifold. 
 
On the one hand, the present approximate solution of the BFKL equation ((1.5-9)
with the parameters listed in Table 2) can 
be used in the future to compare it with the numerical 
solutions of this equation as soon as they will be available (and there is a 
strong need for such numerical solutions~!).

Further, our interpolating formula may clarify the still open question:
what is the Pomeron~? While the answer is known within the context of 
the analytic $S$-matrix theory (that gave rise to the notion of the 
"Pomeron"), namely that 
it is an isolated moving $j$-plane singularity with vacuum quantum numbers and 
with $\alpha(0) = 1$, an alternative definition may arise from QCD, namely 
that the Pomeron is the solution of the BFKL equation. In that case the 
deviation from the simple $s^{\alpha(t,Q^2)}$ behavior may be indicative of 
the difference between the two. The model presented in this paper shows it.

Another aspect of the Pomeron studies is its "hardness" at low (or 
vanishing) $Q^2$. It is well known (see \eg\cite{jenk}) that in the present 
energy range, the cross sections (or structure functions) can
well be described by logarithmic functions as well (rather than powers in $s$
(or $1/x$)). Apart from numerical fits, this phenomenon has also a physical 
interpretation~\cite{jkp}: in the presently available energy range there is 
sufficient phase space available only for a finite (small) number of gluons 
rungs in the Pomeron ladder, each contributing with a power in 
$\ell n s \ $ (or $\ell n(1/x))$, resulting thus in a 
$\ell n(s), \ell n^2(s),...$ behavior of 
the cross sections (same for the SF). As repeatedly stressed, such logarithmic 
parametrizations, being equally efficient phenomenologically, have the 
advantage of being consistent with the unitarity bounds. Note that the 
extrapolation between the low $Q^2$ logarithmic behavior and high $Q^2$
asymptotics may be more complicated than that presented above. For that 
purpose the model of \cite{bgp95}, combining the logarithmic behavior at 
$Q^2=0$ with a "supercritical" power behavior far off shall may be appropriate.

Finally, we remind that the "BFKL-Pomeron" 
presented in the present paper is not a direct solution of the 
BFKL equation. Instead, it interpolates betwen the two known asymptotic 
solutions of that equation and as such is a small step forwards with respect 
to those previously known. Our poor knowledge of other solutions (in 
different kinematical regions~?) of the BFKL equation shows on one hand the 
complexity of the problem, but on the other hand it is clear that further
progress in theory cannot be achieved without a better understanding of 
such basic objects in high energy physics as the bound state of two gluons or
the elastic scattering amplitude.  

\medskip
{\bf Acknowlegdments.}
We thank V. Fadin, E. Kuraev and L. Lipatov for numerous discussions on the 
Pomeron.

                         \vfill\eject

\begin{displaymath}
\begin{array}{|c|c|c|c|}
\hline
{\rm Observable} \ &{\rm Number\, of \,points}&\chi^2  &\chi^2\\
{\rm Experiment-year\ of\ pub.,\ Ref} & &
  &\cite{djp98} \\
\hline 
 F_2^p                 &            &          &  \\
H1-95      \ \ \cite{ah95}  &  93\ (0\ {\rm in}\ \cite {djp98}) &77  & - \\
H1-96      \ \ \cite{ai96}  & 193 &149 & 110 \\
H1-97      \ \ \cite{ad97}  &  44 &55 & 20 \\
ZEUS-96    \ \ \cite{de96}  & 188\ (0\ {\rm in}\ \cite {djp98}) &256 & - \\
ZEUS-97    \ \ \cite{br97}  &  34\ (0\ {\rm in}\ \cite {djp98}) &13  & - \\
ZEUS-98    \ \ \cite{br98}  &  44\ (0\ {\rm in}\ \cite {djp98}) &28  & - \\
E665-96    \ \ \cite{ad96}  &  91\ (0\ {\rm in}\ \cite {djp98}) &101 & - \\
NMC-95     \ \ \cite{ar95}  & 156\ (0\ {\rm in}\ \cite {djp98}) &177 & - \\
SLAC-90/92 \ \ \cite{slac}  & 136\ (0\ {\rm in}\ \cite {djp98}) &145 & - \\ 
BCDMS-89   \ \ \cite{be89}  & 175\ (0\ {\rm in}\ \cite {djp98}) &257 & - \\
\hline
\sigma_{tot}^{\gamma ,p} \ \ \cite{ca75} & 99\ (73\ {\rm in}\ \cite {djp98})
  &177   & 73  \\
\hline
{\rm Total}             & 1253\ (310\ {\rm in}\ \cite {djp98})
   &1434  & 203  \\
\chi^2\ / \ d.o.f.&     - &1.16   & 0.67  \\
\hline
\end{array}                                                      
\end{displaymath}

\centerline {\bf Table 1.}            
\medskip
Observables used in the fitting procedure. The  
complete available experimental kinematical range is taken into account for 
$Q^2$ (\ie $0.\leq Q^2$ (\g2) $ \leq 5000$,
$x$ ($2.\ 10^{-6}\leq x\leq 0.75$),
with the lower limit $W\geq 3$ GeV. 

Also shown is the distribution of the partial
$\chi^2$ for each subset of data used in our fit with the 
parameters listed in Table 2. The results from \cite{djp98} are also shown.

\bigskip
\def\init{\tabskip 0pt\offinterlineskip}
\def\crr{\cr\noalign{\hrule}}
$$\vbox{\init\halign to 6 cm{
\strut#&\vrule#\tabskip=1em plus 2em&                                          
\hfil$#$\hfil&
\vrule#\tabskip 0pt\crr
&&{\bf low\ }x                        &\crr
&&{\bf A}= .1365          &\cr
&&{\bf a{\rm \ (GeV)}^2}=.2372  &\cr
&&{\bf \gamma_2}=.02158   &\cr
&&{\bf Q_0^2{\rm \ (GeV)}^2}=.1991   &\cr
&&{\bf Q_1^2{\rm \ (GeV)}^2}=1490.   &\cr
&&{\bf B}=1.944 &\cr
&&{\bf b{\rm \ (GeV)}^2}=1.804 &\cr
&&{\bf \alpha_r}=.3603 &\crr
&&{\bf large\ }x                       &\crr
&&{\bf Q^2_p{\rm \ (GeV)}^2}=.1864     &\cr 
&&{\bf p_0}=-26.32 &\cr
&&{\bf p_\infty}=10.52 &\cr   
&&{\bf Q^2_r{\rm \ (GeV)}^2}=13.86 &\cr
&&{\bf r_0}=2.601 &\cr   
&&{\bf r_\infty}=3.846 &\crr
}}$$

\centerline {\bf Table 2.}

Free parameters used in our fit.
The non-fitted values are $ \epsilon=0.08$ (fixed from \cite{eps}) and  
$\gamma_1 = 2.4$ (suggested by QCD).

\newpage

\begin{center}
\epsfig{figure=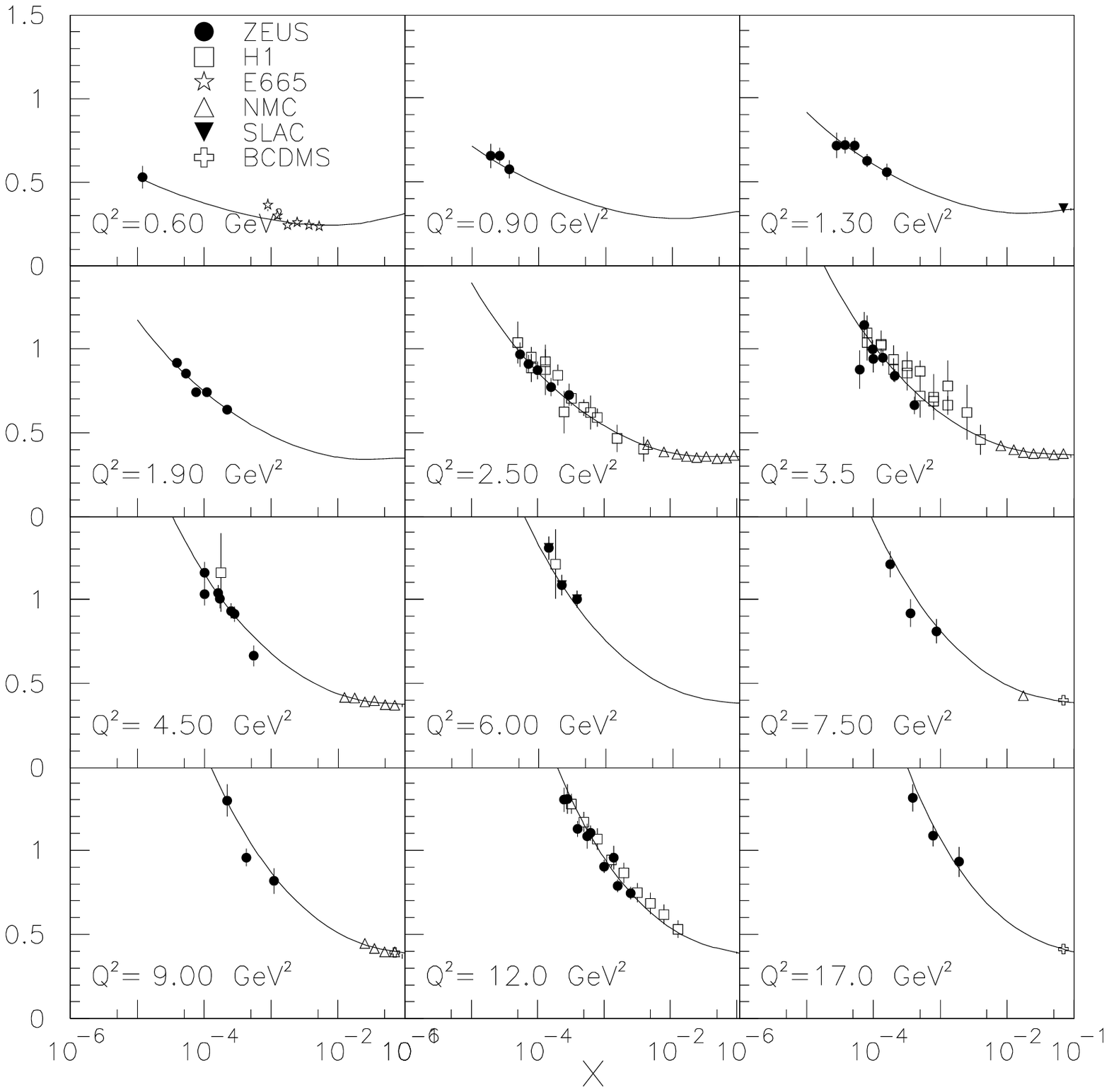,width=16cm}
\end{center}
\bigskip
{\bf Fig.~1}
Proton structure function $F_{2}(x,Q^2)$ as a function of $x$ at low $Q^2$.
The data shown are listed in Table 1,
the error bars represent the statistical and systematic errors added in
quadrature, the solid curves are the results of our fit (the parameters are 
listed in Table 2).

\begin{center}
\epsfig{figure=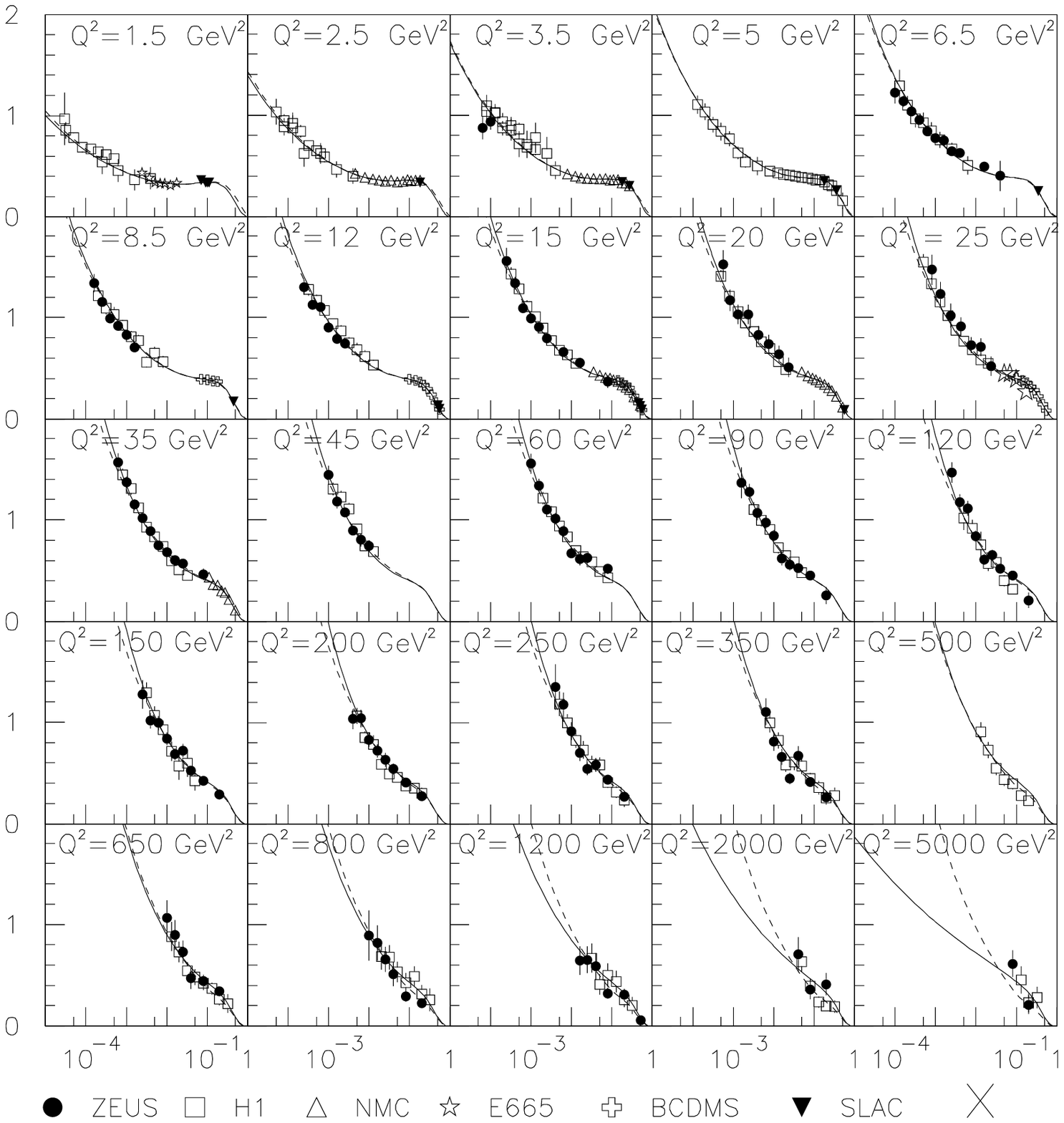,width=16cm}
\end{center}
\bigskip
{\bf Fig.~2}
Proton structure function $F_{2}(x,Q^2)$ as a function of $x$  at
various values of $Q^2$ (see also Fig.~1). The dashed lines are the results of
the ALLM model \cite{allm}.


\begin{center}
\epsfig{figure=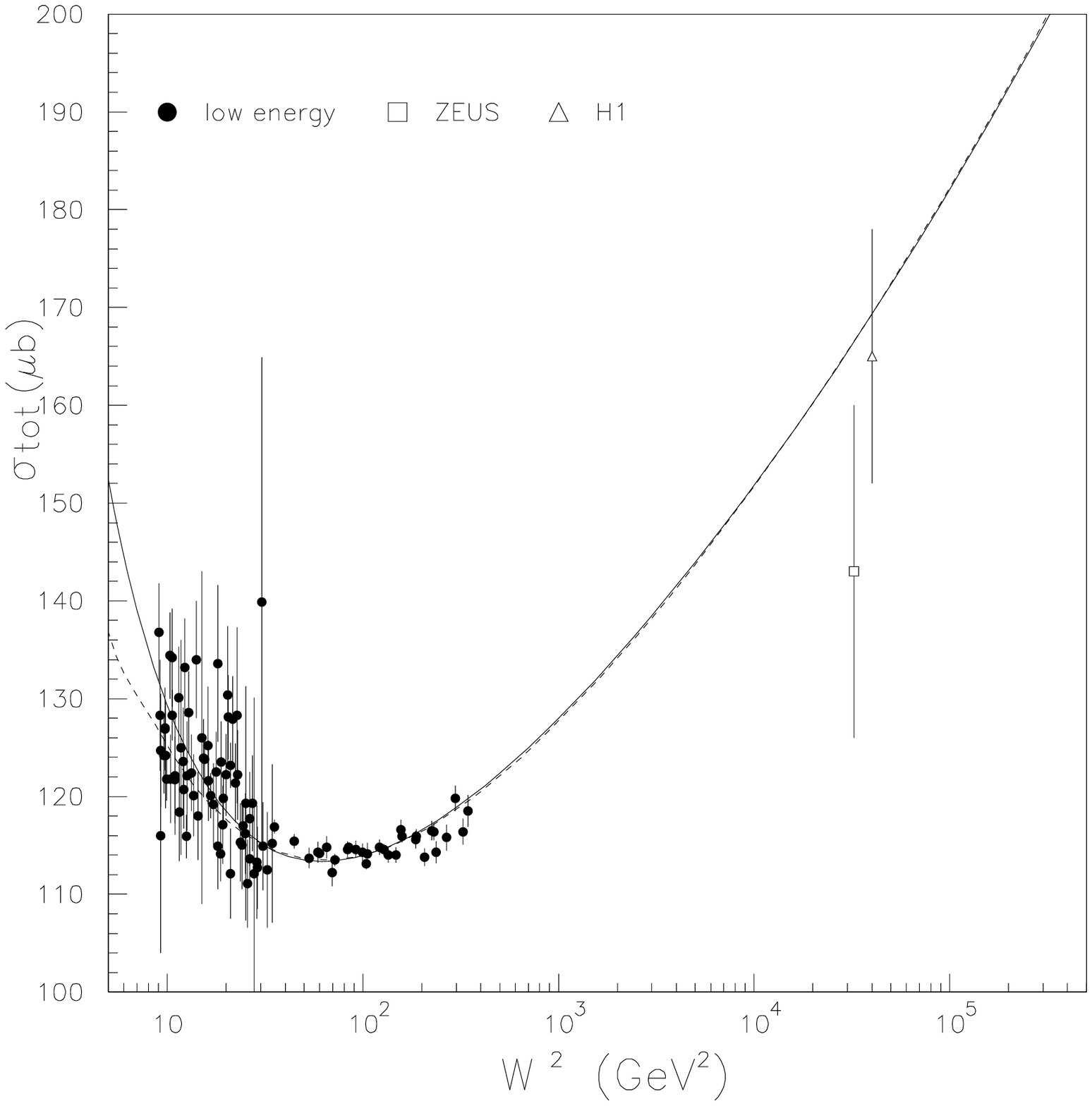,width=12cm}
\end{center}
\bigskip
{\bf Fig.~3.}
Limiting case to the real photon-proton total cross-section 
$\sigma_{tot}^{(\gamma,p)}$ as a function of $W^2$, square center of mass 
energy (see also Fig.~1). The dashed lines are calculations from the ALLM 
model \cite{allm}.

\begin{center}
\epsfig{figure=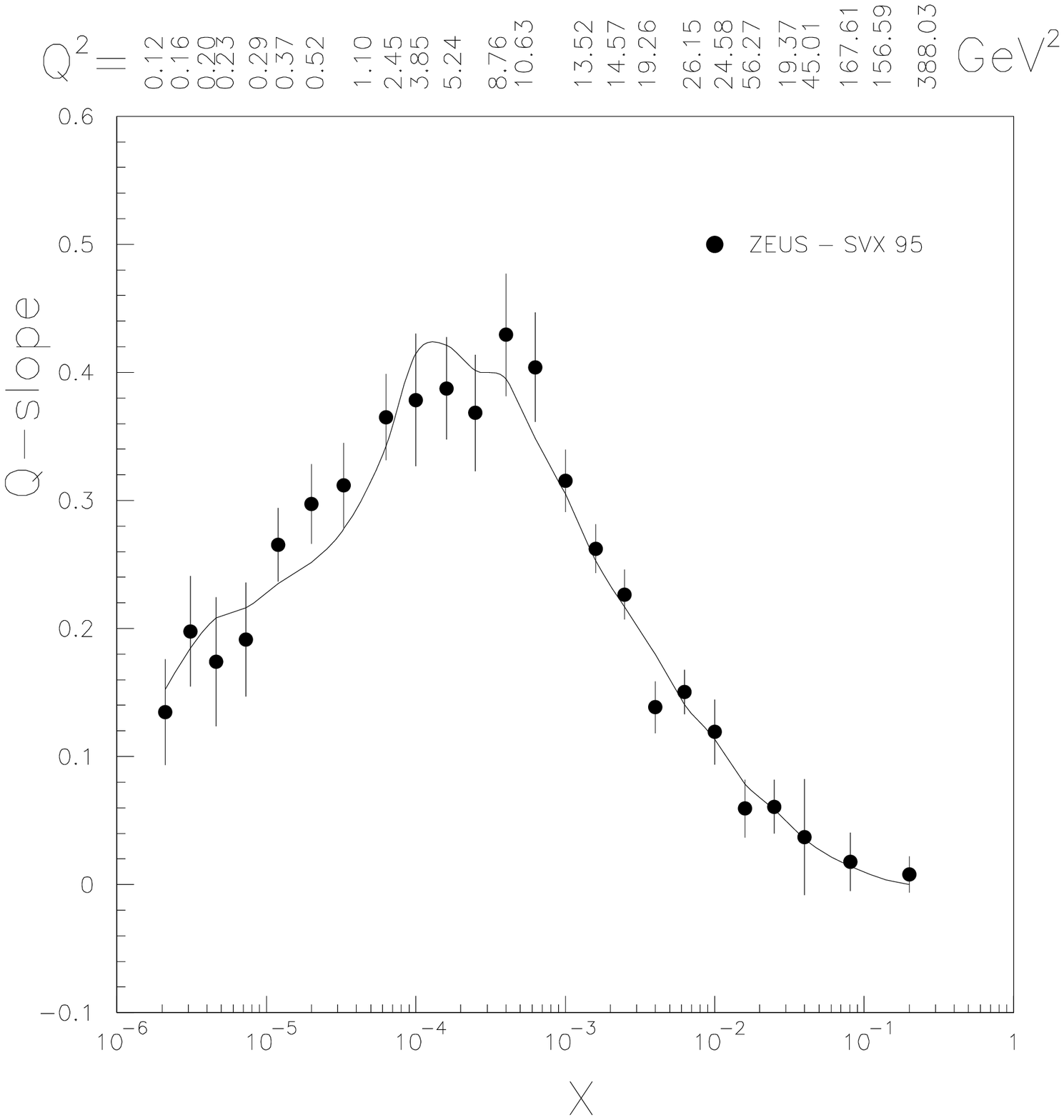,width=12cm}
\end{center}
\bigskip
{\bf Fig.~4.}
Q-slope $={\partial F_2(x,Q^2)\over\partial\ell n Q^2}$ as 
a function of $x$ (for the indicated $Q^2$ values).
The experimental points are from \cite{br98}. The predictions in 
the present model are given in continuous line.

\begin{center}
\epsfig{figure=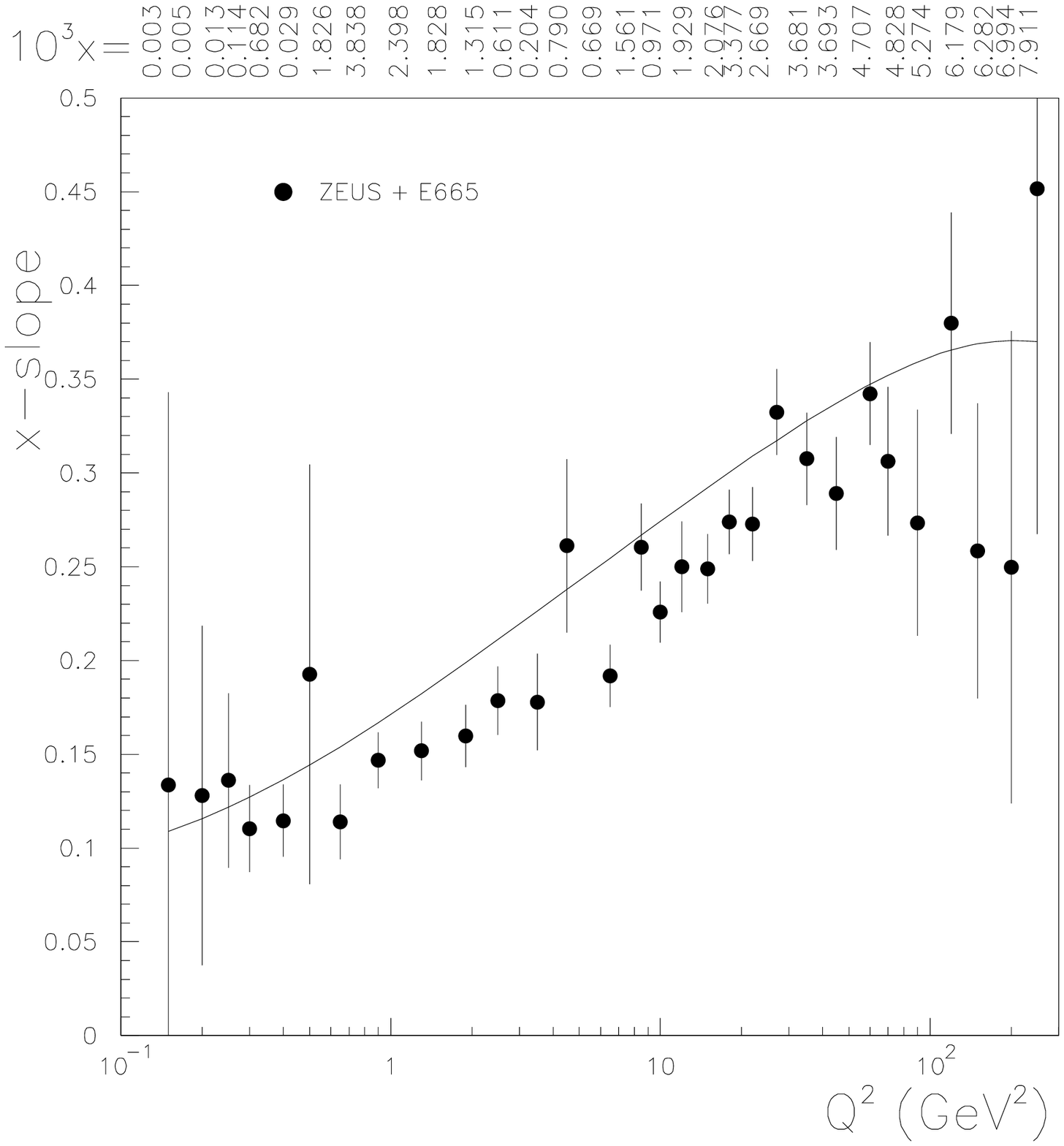,width=12cm}
\end{center}
\bigskip
{\bf Fig.~5.}
$\lambda_{eff}(<x>,Q^2)$ experimental points from \cite{br98} as 
a function of $Q^2$ (for the indicated $<x>$ values), compared with the 
calculated $x-$ slope values 
$={\partial\ell n F_{2}(x,Q^2)\over\partial(\ell n (1/x))}$. The predictions in 
the present model are given in the approximation including only the low $x$ 
Pomeron contribution (2.5).

\begin{center}
\epsfig{figure=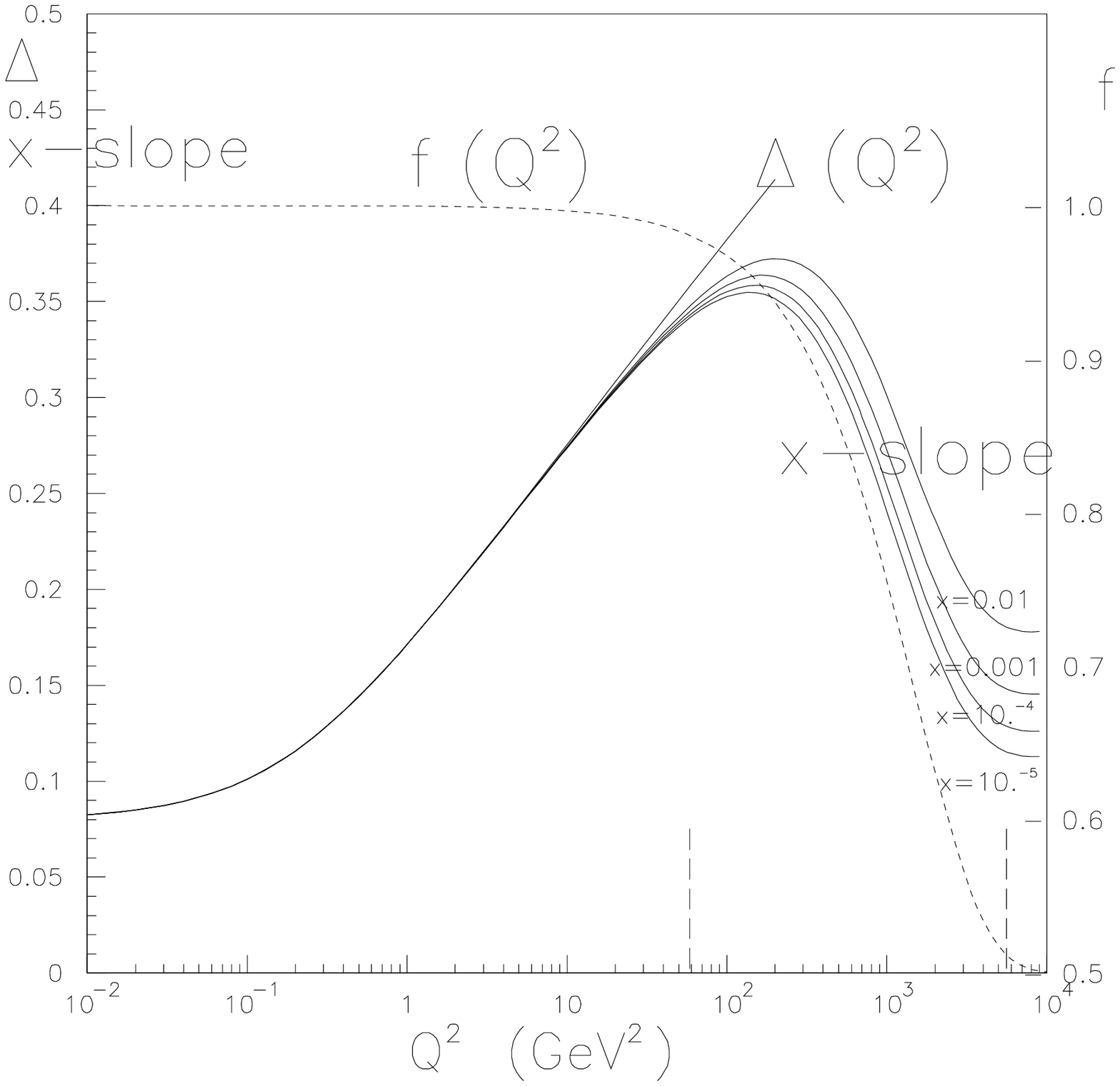,width=12cm}
\end{center}
\bigskip
{\bf Fig.~6.}
$x-$slope $={\partial\ell n F_{2}(x,Q^2)\over\partial(\ell n (1/x))}$
as a function of $Q^2$ (for the indicated $x$ values).
The predictions in the present model (solid line) are calculated in 
the low $x$ Pomeron approximation (see Fig.~5). On the same left 
scale the
exponent $\tilde\Delta$ (2.8), equivalent to the Pomeron intercept-1 when
$f(Q^2)\approx 1$ is also plotted. This function $f(Q^2)$ is also shown in dash line (right
scale). The estimated transition region is between the vertical landmarks (see
the text). 

\end{document}